\pdfoutput=1
\documentclass[conference]{IEEEtran}

\usepackage{url}

\usepackage{graphicx} 
\usepackage{balance}  
\usepackage{booktabs} 
\usepackage{ccicons}  
\usepackage{ragged2e} 

\usepackage{xcolor}

\IEEEoverridecommandlockouts
\title{AI Development for the Public Interest: From Abstraction Traps to Sociotechnical Risks}

\author{
\thanks{Authors arranged alphabetically to indicate equal contribution.}
\IEEEauthorblockN{McKane Andrus}
\IEEEauthorblockA{Partnership on AI \\
mckane@partnershiponai.org}
\and
\IEEEauthorblockN{Sarah Dean}
\IEEEauthorblockA{Electrical Engineering and Computer Sciences\\ UC Berkeley \\
dean\_sarah@berkeley.edu}
\and
\IEEEauthorblockN{Thomas Krendl Gilbert}
\IEEEauthorblockA{Center for Human-Compatible AI \\ UC Berkeley \\
tg340@berkeley.edu}
\and
\IEEEauthorblockN{Nathan Lambert}
\IEEEauthorblockA{Electrical Engineering and Computer Sciences\\ UC Berkeley \\
nol@berkeley.edu }
\and
\IEEEauthorblockN{Tom Zick}
\IEEEauthorblockA{Berkman Klein Center for Internet and Society \\ Harvard University \\
tzick@cyber.harvard.edu}
}

\author{\IEEEauthorblockN{McKane Andrus\IEEEauthorrefmark{1},
Sarah Dean\IEEEauthorrefmark{2},
Thomas Krendl Gilbert\IEEEauthorrefmark{3}, 
Nathan Lambert\IEEEauthorrefmark{2} and
Tom Zick\IEEEauthorrefmark{4}}
\IEEEauthorblockA{\textit{Authors arranged alphabetically. }\IEEEauthorrefmark{1}Partnership on AI, San Francisco, CA.\\
}
\IEEEauthorblockA{\IEEEauthorrefmark{2}Department of Electrical Engineering and Computer Sciences, 
University of California, Berkeley. \\
}
\IEEEauthorblockA{\IEEEauthorrefmark{3}Center for Human-Compatible AI, 
University of California, Berkeley. \\
}
\IEEEauthorblockA{\IEEEauthorrefmark{4}Berkman Klein Center for Internet and Society, Harvard University. \\
\small{ mckane@partnershiponai.org, \{dean\_sarah, tg340, nol\}@berkeley.edu, tzick@cyber.harvard.edu}
}}

\begin{document}

\maketitle

\begin{abstract}
Despite interest in communicating ethical problems and social contexts within the undergraduate curriculum to advance Public Interest Technology (PIT) goals, interventions at the graduate level remain largely unexplored. This may be due to the conflicting ways through which distinct Artificial Intelligence (AI) research tracks conceive of their interface with social contexts. In this paper we track the historical emergence of sociotechnical inquiry in three distinct subfields of AI research: AI Safety, Fair Machine Learning (Fair ML) and Human-In-the-Loop (HIL) Autonomy. We show that for each subfield, perceptions of PIT stem from the particular dangers faced by past integration of technical systems within a normative social order. We further interrogate how these histories dictate the response of each subfield to conceptual traps, as defined in the Science and Technology Studies literature. Finally, through a comparative analysis of these currently siloed fields, we present a roadmap for a unified approach to sociotechnical graduate pedogogy in AI.

\end{abstract}


%
\IEEEpeerreviewmaketitle

\section{Introduction}

Recent years have seen an increasing public awareness of the profound implications of widespread artificial intelligence (AI) and large scale data collection. It is now common for both large tech companies and academic researchers to motivate their work on AI as interfacing with the ``public interest," matching external scrutiny with new technical approaches to making systems fair, secure, or provably beneficial. However, developing systems in the public interest requires researchers and designers to confront what has been elsewhere referred to as the “sociotechnical gap,” or the divide between the intended social outcomes of a system and what is actually achieved through technical methods \cite{ackermanIntellectual2000}.

Interventions in Computer Science (CS) education have made strides towards providing students with frameworks within which to evaluate technical systems in social contexts \cite{integrated, integrated2}. These curricular modifications have drawn on fields like Law, Philosophy, and Science and Technology Studies  (STS) to create both dedicated and integrated coursework promoting human contexts and ethics in CS \cite{syllabi}. However, as the majority of these courses are currently offered at the undergraduate level, graduate students may not reap the benefits of such reforms \cite{syllabi}. Given the role of graduate students as not only teachers, but drivers of cutting edge research and future decision makers in industry and academia, interventions aimed at them may play an outsized role in forwarding PIT goals.  
\begin{figure}[t]
    \centering
    \includegraphics[width=.8\columnwidth]{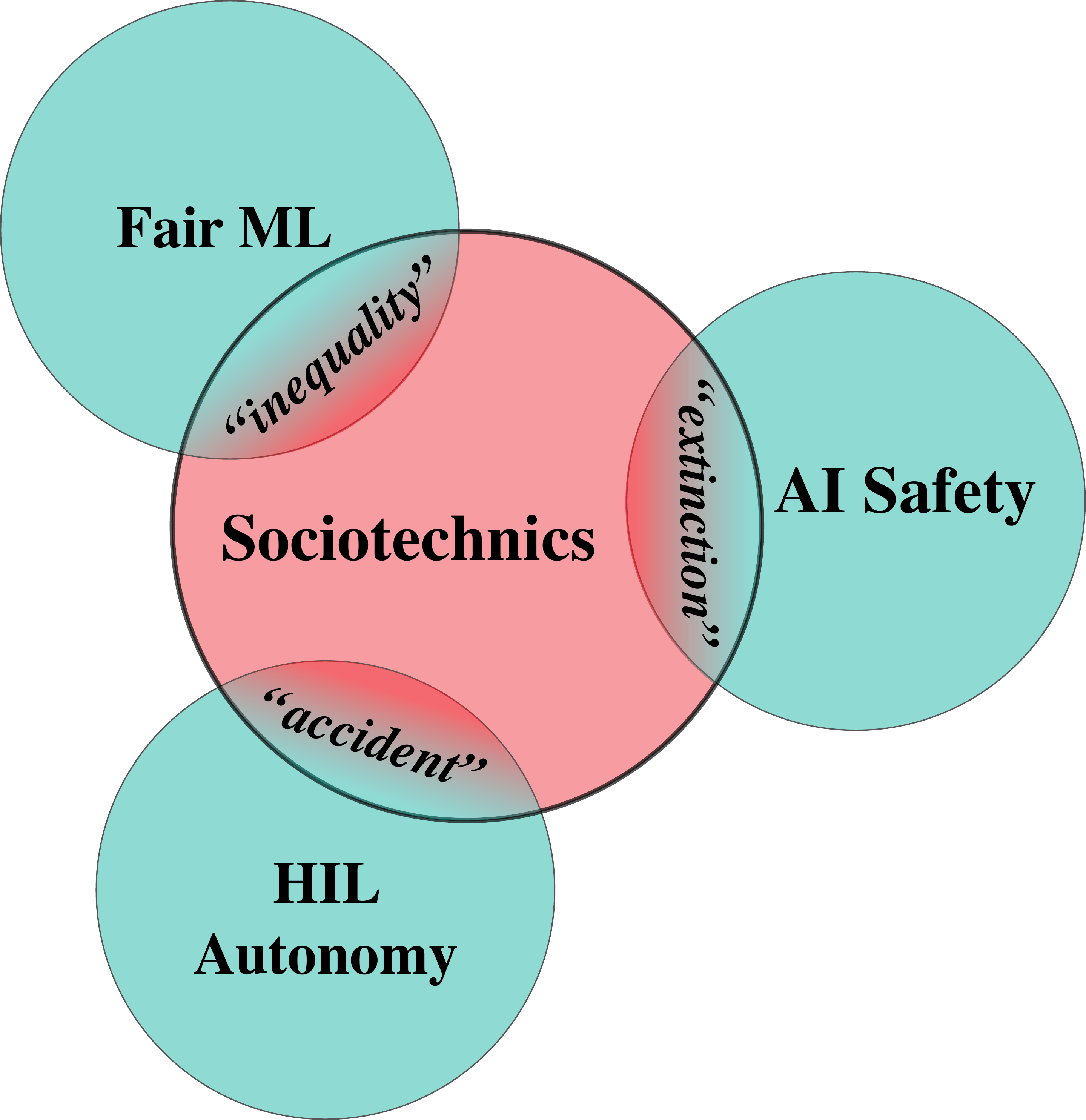}
    \caption{Three contemporary areas of artificial intelligence research whose overlapping forms of sociotechnical inquiry remain problematically defined: AI Safety, Fair Machine Learning, and Human-in-the-Loop Autonomy.}
    \label{fig:concept}
\end{figure}

It is challenging to pin down what it would mean to train a graduate AI researcher to address the sociotechnical gap.
A key source of tension is the place of the sociotechnical in AI development: while practitioners claim to be working on technical solutions to social problems, theoretical and methodological formulations of the sociotechnical are inconsistent across prominent AI subfields, making it unclear if current initiatives in pedagogy and research are advancing, undermining, or neglecting the public interest.

In this paper, we go beyond coursework to analyze the historical and technical shifts behind the current conception of sociotechincal risks in prominent AI subfields.
We look to existing research domains that grapple with the social and technical spheres at distinct levels of abstraction, and examine how their limitations and insights reflect nascent, if problematic, \textit{forms of inquiry} into the sociotechnical. We assess current research in the socially-oriented subfields of AI highlighted in Fig.~\ref{fig:concept}, namely AI Safety; Fairness in Machine Learning (Fair ML); and Human-in-the-Loop (HIL) Autonomy.
AI Safety focuses on value-alignment of future systems and cautions against developing AI systems fully integrated with and in control of society.
Fair ML works to reduce bias 
in algorithms with potentially deleterious effects on individuals or groups.
HIL Autonomy is a term we use to encompass emerging work on both human-robot interaction and cyber-physical systems. These research areas explore how to optimize interactions of autonomous systems with human intentions in the loop.
By tracing the history of these subfields in a comparative fashion, we are able to characterize their distinct orientations towards sociotechnical challenges, highlighting both insights and blindspots. The goal of this analysis is not to capture each subfield's research agenda exhaustively, which is far beyond our scope. Instead, it is to highlight how these agendas claim relative access to some feature of the sociotechnical in the way they represent social problems as technically tractable.
In doing so, they claim legitimacy and authority with respect to public problems.

We next refine this comparative history with lenses borrowed from the Science and Technology Studies (STS) literature, emphasizing the ways the subfields interface with sociotechnical risks. 
In particular, we portray how certain risks are deferred within each subfield's agenda.
This deferral often takes the form of skillfully avoiding ``abstraction traps" that have been recently highlighted by \cite{selbst2019fairness}.
Such avoidance is important from a technical standpoint.
However, true engagement with the sociotechnical requires reflexively revealing and resolving risks beyond the piecemeal formalisms that have defined each subfield's historical trajectory.
We conclude with a brief sketch of pedagogical interventions towards this goal. Beyond classroom ethics curricula, we propose an agenda for clinical engagement with problems in the public interest as a part of graduate training in AI. 
This training would inculcate a direct appreciation of sociotechnical inquiry in parallel with the acquisition of specific technical skillsets.
It would prepare practitioners to evaluate their own toolkits discursively, rather than just mathematically or computationally.

\section{Emerging Sociotechnical Subfields of AI: Safety, Fairness, Resiliency}\label{sec:subfields}

Recent technical work grappling with the societal implications of AI includes developing provably safe and beneficial artificial intelligence (AI Safety), mitigating classification harms for vulnerable populations through fair machine learning (Fair ML), and designing resilient autonomy in robotics and cyber-physical systems (HIL Autonomy). 
At present, these areas constitute heterogeneous technical subfields with substantive overlaps, but lack discursive engagement and cross-pollination.

Below, we outline the history and motivating concerns of each subfield and identify key developments and convenings. 
We highlight how the technical research agendas stem from distinctly sociotechnical concerns and will require interdisciplinary engagement
to fully map out their stakes. 
By placing these subfields' agendas in 
the context of the local sociotechnical risks (respectively \textit{extinction}, \textit{inequity}, and \textit{accident}), we argue that the representative technical formations (\textit{forecasting} and \textit{value alignment}, \textit{fairness criteria} and \textit{accountability}, \textit{controllability} and \textit{reachability}) iterate on those risks without reflexively interrogating or normatively addressing them.
Throughout this work we use the term AI system to refer to a technical system with significant automated components (e.g. automated decision making systems or self driving cars).
We note that AI is also a distinct field of academic study, further discussed in this section.

\subsection{AI Safety}
The field of artificial intelligence (AI) has often situated itself within a wider disciplinary context. 
Famously, at the foundational summits at Dartmouth and MIT, computer scientists, logicians, and psychologists came together to chart a course for artificial intelligence 
to arrive at human-like capabilities~\cite{russell2002artificial}. 
That course, which laid the groundwork for 
``good-old-fashioned" AI, had an incredible number of hiccups and was eventually overwritten. A similarly diverse group of disciplinary representatives moved the field away from symbolic and comprehensive logical reasoning to either more situated, interactionist understandings of cognition~ \cite{dreyfus1992computers,agre1997computation} or to biologically-inspired, connectionist strategies of learning~\cite{sun2014connectionism}.
Following these interdisciplinary developments, there has been a growing concern about the capabilities of AI systems to endanger humans and society writ large~\cite{bostrom2003ethical,yudkowsky2008artificial,armstrong2016racing}. 
Motivated both by longstanding concerns about the possibility of a ``technological singularity"~\cite{kurzweil2005singularity} as well as recent expansive applications of machine learning in critical infrastructure domains, many AI Safety promoters fear that AI researchers are approaching a level of capability that will expand beyond their control~\cite{kurzweil2005singularity, bostrom2017superintelligence}. 

Belief in the prospect of an arbitrarily capable intelligent agent beyond designer control has raised the prospect of \textit{extinction}, whether of humanity or all natural life, as a clear and present danger for AI development, serving as this field's distinct sociotechnical risk scenario.
Regardless of the likelihood of such a scenario, the nascent field of AI Safety has arisen to preemptively confront these dangers. 
AI Safety takes a rather radical approach to the type of systems level thinking that we discuss, often viewing technical developments on a much longer and wider timescale---see, for example, the published work on AI arms races and potential research agendas to avert them \cite{armstrong2016racing,ramamoorthy2018beyond}.  As such, a common feature of AI Safety research is \textit{forecasting} future AI capabilities against various time horizons.

Despite recent high-profile endorsements from computer scientists and philosophers such as Stuart Russell \cite{russell2019human} and Nick Bostrom \cite{bostrom2017superintelligence}, AI Safety is still a nascent research community. At present there is no independent conference for this field, although workshops and panels on AI Safety have become a regular fixture of larger AI venues such as NeurIPS, ICML, and AAAI, while specific AI-safety oriented research labs (e.g. CHAI, OpenAI) host invited technical presentations on a semiweekly or monthly basis. The field has also attracted interest from research centers and philanthropic organizations dedicated to the study and mitigation of long-term existential risk, as well as industry leaders in AI.

Central to this work is a motivation to align AI systems with {intended} rather than {specified} rewards, as humans struggle to make explicit the rich normative context of their own goals and behaviors. Through this shift, AI Safety adjusts the framing of classical AI development towards ``provably beneficial" rather than merely optimal systems.
Under this  framing, researchers focus largely on the problem of \emph{value alignment}, i.e. whether or not an AI agent's programmed objective matches those of relevant humans or humanity as a whole~\cite{soares2017agent}.
For example, by understanding the problem of aligning an AI agent with a human collaborator as a problem of inverse reinforcement learning, researchers seek to solve this issue with a largely technical approach by borrowing core principles from economic game theory~\cite{hadfield2016cooperative}.

Considered as a whole, extended sociotechnical inquiry in {AI Safety} remains limited to catastrophic risk evaluation in cases where humanity's survival is at stake---a scale of concern that is not often found in engineering disciplines. Moreover, rigorous formal work often relies on intuition from mechanism design (e.g. an objectives-first approach, perfectly rational agents) whose assumptions inherit some of the formal limitations of and controversies surrounding prospect theory and social choice theory.
Stemming from AI Safety, we see vigorous discussions surrounding AI Policy \cite{brundage2020toward}, ethics \cite{gabriel2020artificial}, and even reflexive interrogations as a practice in forecasting  \cite{grace2018will}. While lacking some qualities of sociotechnical inquiry, in particular a deeply reflexive methodology and historical orientation, we see potential to pivot these discussions away from narrowly-framed thought experiments about paperclip-maximizing robots \cite{yudkowsky2008artificial} towards comparative investigations of the normative stakes of distinct AI-society interfaces.


\subsection{Fairness in Machine Learning} 
The field of machine learning (ML) emerged in the late 1950s with the design of a self-improving program for playing checkers~\cite{samuel1959some} and quickly found success with static tasks in {pattern classification}, including applications like handwriting recognition~\cite{jnilsson1965learning}.
ML techniques work by detecting and exploiting statistical correlations in data, towards increasing some measure of performance.
A prominent early machine learning algorithm was the perceptron~\cite{rosenblatt1957perceptron}, an example of supervised classification, perhaps the most prevalent form of ML.
In this setting, a classifier (or model) is trained with labelled examples, and its performance is measured by its accuracy in labelling new instances.
The perceptron spurred the development of deep learning techniques mid-century~\cite{olazaran1996sociological}; however, they soon
fell out of favor, only having great success in recent decades in the form of neural networks via the increasing availability of computation and data.
Many ML algorithms require large datasets for good performance, tying the field closely with ``big data.''
However, optimizing predictive accuracy does not generally ensure beneficial outcomes when predictions are used to make decisions, a problem that becomes stark when individuals are harmed by the classification of an ML system. 

The \textit{inequality} resulting from system classifications is the central sociotechnical risk of concern to practitioners in this the subfield of Fair ML.
A growing awareness of the possibility for bias in data-driven systems developed over the past fifteen years, starting in the data mining community~\cite{pedreshi2008discrimination} and echoing older concerns of bias in computer systems~\cite{friedman1996bias}.
The resulting interest in ensuring ``fairness'' was
further catalyzed by high profile civil society investigation (e.g. ProPublica's Machine Bias study, which highlighted racial inequalities in the use of ML in pretrial detention) and legal arguments that such systems could violate anti-discrimination law~\cite{barocas2016big}. 
At the same time, researchers began to investigate model ``explainability'' in light of procedural concerns around the black box nature of deep neural networks.
The research community around Fairness in ML began to crystallize with the ICML workshop on Fairness, Accountability, and Transparency in ML (FAT/ML), and has since grown into the ACM conference on Fairness, Accountability, and Transparency (FAccT) established in 2017.

By shifting the focus to fairness properties of learned models, Fair ML adjusts the framing of the ML pipeline away from a single metric of performance.
There are broadly two approaches: individual fairness, which is concerned with similar people receiving similar treatment~\cite{dwork_fairness_2012}, and group fairness which focuses on group parity in acceptance or error rates~\cite{barocas-hardt-narayanan}.
The details of defining and choosing among these \emph{fairness criteria} amount to normative judgements about which biases must be mitigated, with some criteria being impossible to satisfy simultaneously.
Much technical work in this area focuses on algorithmic methods for achieving fairness criteria through either pre-processing on the input data~\cite{calmon2017optimized}, in-processing on the model parameters during training~\cite{zafar2019fairness}, or post-processing on model outputs~\cite{hardt2016equality}.

The Fair ML community is oriented towards the sociotechnical, engaging actively with critiques from STS perspectives.
FAccT is a strong locus of interdisciplinary thought within computer science,
and the addition of \emph{transparency} and \emph{accountability} to the title opens the door to a wider range of interventions.
Building upon model-focused concepts like explainability, blendings of technical and legal concepts of recourse~\cite{ustun2019actionable} and contestability~\cite{mulligan2019shaping} widen the frame to explicitly consider the reaction of individuals to their classification. Similarly, there have been multiple calls to re-center 
stakeholders
to understand how explanations are interpreted and if they are even serving their intended purpose \cite{miller2019explanation,bhatt2020machine}.
The community is increasingly open to discussing scenarios in which technical intervention, like the police use of facial recognition, is not desired. This encompasses both
technical resistance~\cite{kulynych2020pots} and procedural approaches to delineating the valid uses of data~\cite{gebru2018datasheets} and models~\cite{mitchell2019model}.

\subsection{Human-in-the-Loop Autonomy}

As many of the earliest robotic systems were remotely operated by technicians, the field of robotics has always had problems of human-robot interaction (HRI) at its core~\cite{goodrich2008human}.
Early work was closely related to the study of human factors, an interdisciplinary endeavor drawing on engineering psychology, ergonomics, and accident analysis~\cite{bainbridge1983ironies}.
With advancements in robotic capabilities and increasing autonomy, the interaction paradigm grew beyond just teleoperation to \emph{supervisory control}.
HRI emerged as a distinct multidisciplinary field in the 1990s with the establishment of the IEEE International Symposium on Robot \& Human Interactive Communication.
Modern work in this area includes modeling interaction from the perspective of the autonomous agent (i.e. robot) rather than just the human overseer. 
By incorporating principles from the social sciences and cognitive psychology, HRI uses predictions and models of human behavior to optimize and plan. This work mitigates the sociotechnical risk of \textit{accidents} -- defined specifically as states in which physical difficulties or mishaps occur.
Such physical risks are mitigated by making models robust to these potentially-dangerous conditions.

Digital technology has advanced to the point that many systems are endowed with autonomy beyond the traditional notion of a robotic agent, including traffic signal networks at the power grid. 
We thus consider the subfield of \emph{HIL Autonomy} to be the cutting edge research that incorporates human behaviors into robotics and cyber-physical systems.
This subfield proceeds in two directions: 1)  innovations in physical interactions via sensing and behavior prediction; 2) designing for system resiliency in the context of complicated or unstable environments. 
These boundaries are blurring in the face of increasingly computational methods and the prospective market penetration of new technologies. For example, the design of automated vehicles (AVs) poses challenges along many fronts.
For more fluent and adaptable behaviors like merging, algorithmic HRI attempts to formalize models for one-on-one interactions.
At the same time, AVs pose the risk physical harm, so further lines of work integrate these human models to ensure safety despite the possibility of difficult-to-predict actions. 
Finally, population-level effects (e.g. AV routing on traffic throughput and induced demand) require deeper investigation into interaction with the social layer.

The emerging subfield of HIL Autonomy uses ideas from classical \textit{control theory} while trying to quantify and capture the risk and uncertainty of working with humans~\cite{baheti2011cyber,banerjee2011ensuring}. 
It thus inherits some of the culture around verifying safety and robustness through a combination of mathematical tools and physical redundancy,
due to a history of safety-critical applications in domains like aerospace.
Technical work in this area typically entails including the human as part of an under-actuated dynamical system \cite{sadigh2017active, wu2018stabilizing}, such as a un-modeled disturbance. Through this lens, human-induced uncertainty is mitigated by predicting behavior in a structured manner, maintaining the safety of the system through mathematical robustness guarantees ~\cite{bajcsy2019scalable}. 
To make this concrete, a lane-change maneuver in an AV might include both an aggressive driving plan that takes likely human behaviors into account as well as a \textit{reachability} safety criterion which could be activated via feedback if observed human behavior falls outside of the expected distribution.
At a higher level of planning, the lane change maneuver may only be directed if it is expected to be advantageous for global traffic patterns.

The extent to which HIL Autonomy engages with the sociotechnical is thus far limited. 
Human-centered research focuses on localized one-to-one interactions, while research considering more global interactions remains largely in the realm of the technical.
However, the critical ``alt.HRI" track at the ACM/IEEE International Conference on Human-Robot Interaction indicates an emerging interest in how robotic systems interact with society more broadly.
In such venues, questions are raised surrounding how robots interact with social constructions of race \cite{bartneck2018robots,sparrow2020robotics} and issues of robot-community integration are being studied in settings ranging from healthcare~\cite{herath2020arts+} to gardening~\cite{verne2020adapting}.
There is also work which considers the incorporation of social values into cyber-physical systems, e.g. fair electricity pricing for smart  grids~\cite{javed2019fairness}.
While our identification of this emerging subfield is perhaps more speculative than the previous two, the physical realization of AI technologies will remain a crucial site of sociotechnical inquiry.

\section{Sociotechnical Integration}

While the subfields of AI Safety, Fair ML, and HIL Autonomy each consider problems at the interface of technology and human or social factors, there are differences which arise in part to their disparate histories.
One difference is in time-scales.
AI Safety is primarily concerned with long term outcomes of mis-aligned AI development, while Fair ML focuses on practical implementations of individual models and algorithms with imperfect datasets.
HIL Autonomy bisects the two, with both longer term considerations of how numerous autonomous agents will re-define how humans interact in the environment and short term focus on maintaining safety, e.g. in the presence of unexpected adverse road conditions.
Another difference arises from how
the subfields position themselves at different levels of abstraction.
HIL Autonomy is physically grounded, with a history closely tied with embodied interaction with humans and the social layer, while
Fair ML is socially grounded, and has strong instincts for sociotechnical dialog and historical situatedness.
On the other hand, AI Safety positions itself at the highest level of generality, relegating machine learning to the status of a tool and interpreting robotics as an application of formal guarantees.

For these subfields to place their sociotechnical inquiry on firmer foundations, it will be necessary to establish more reflexive relationships with their inherited assumptions about risk. Each subfield interprets itself as filling well-defined sociotechnical gaps, i.e. that there is a discernible divide between social problems and technical agendas. But in fact, the way these subfields have defined and worked on those gaps is itself problematic, piecemeal, and lacking in definition, i.e. it is normatively indeterminate. Reflexive inquiry is needed not to fill those gaps, but to define and interpret them more richly, so that their salience and urgency can be evaluated.

At a minimum, researchers and practitioners must learn to see behind their own technical abstractions to the social reality they assume, recognize that this reality may have been problematically defined, and learn to inquire into these definitions directly, perhaps with the aid of new transdisciplinary tools. 
We now provide a high-level summary of this agenda, moving from a comparison of common \textit{technical traps} to more indeterminate conceptions of sociotechnical risk.

\subsection{Grappling with Shortcomings in Framing}
Each of the subfields discussed in the previous section seeks to expand the technical framing of their parent field to include human and social factors.
In~\cite{selbst2019fairness}, the \emph{framing trap} is introduced as the failure to model the full system of interest (e.g. with respect to a notion of fairness or safety).
Technical researchers are at risk of falling into this trap whenever they draw a \emph{bounding box} around the system that they study.
Often, the consequences of this trap manifest as the \emph{portability trap}~\cite{selbst2019fairness},  which occurs when technical solutions designed for one domain or environment are misapplied to another context.
Technical researchers are at risk of falling into this trap whenever they mistakenly view a bounding box as appropriate to a new context.

The subfields of AI Safety, Fair ML, and HIL Autonomy can be viewed as attempts to avoid the framing trap.
In the fields of AI, ML, and robotics, the workflow often entails \textit{featurization} by defining data or inputs/outputs, \textit{optimization} by fitting a model or designing a control policy, and then \textit{integration} into the larger system.
Researchers in the emerging sub-fields are beginning to understand the downsides of this unidirectional workflow, and the necessity of interrogating the modelling choices made at each step.
For example, AI Safety questions the way that features are used to define optimization objectives in light of potentially catastrophic effects of integration, while Fair ML questions the inequalities arising from model optimization.

Still, sometimes the frame is not opened wide enough.
For example, by failing to account for the larger system in which risk assessments are used, approaches to Fair ML may mistakenly treat loaning decisions the same way they treat pretrial detention, despite salient differences between the financial and criminal justice systems.
By adopting more rigorously a \emph{heterogeneous engineering} approach~\cite{selbst2019fairness}, researchers and practitioners can explicitly determine which properties are not tied to the technical objects under design but to their social contexts.
For example, the aerospace industry is an engineering domain with considerable heterogeneity---an awareness of the regulatory context, from the flight deck procedures to air traffic control, is necessary for the development of flight technologies.

\subsection{Abstraction Traps in AI Research}

To motivate a stronger cross-disciplinary discourse among and outside of these subfields, we now make further use of the framework of abstraction traps provided by~\cite{selbst2019fairness} to point systematically to shortcomings and highlight potential new areas of inquiry.
Alongside the framing and portability traps, we discuss: the formalism trap, the ripple effect trap, and the solutionism trap. 

The \emph{formalism trap} occurs when mathematical formalisms fail to capture important parts of the human context. For example, the fairness of a system is often judged by procedural rather than technical elements, and the perceived reliability may depend more on predictability rather than formally verified safety.
All of the discussed subfields are posed to fall into the formalism trap, which requires a deeper engagement with sociotechnical complexities to avoid.
Ultimately, the validity and desirability of specific metrics arising from mathematical abstractions will be determined through intimate reference to social context rather than technical parsimony.
If systems are not flexible enough to allow for public input, the validity can be compromised.

The \emph{ripple effect trap} occurs when there is a failure to understand how technology affects the social system into which it is inserted. 
AI Safety considers ripple effects to some extent, but in a narrowly formal manner. 
For example, while automated vehicles are known to affect traffic, road, and even infrastructure design, most technical research has focused on incorporating these as features to be modeled rather than questioning the status of AVs as the dominant form of future mobility.
Engagement across the entire sociotechnical stack requires understanding social phenomena like the ``reinforcement politics'' of dominant groups using technology to remain in power and ``reactivity'' like gaming and adversarial behavior. If a system encourages people to behave in an adversarial manner, it may call for utilizing richer design principles to promote cooperation, rather than merely throwing more advanced AI methods at the assumed dynamics.

Finally, the \emph{solutionism trap} occurs when designers mistakenly believe that technical solutions alone can solve complex sociological and political problems. For example, while the legal community has encouraged technical fields to build systems that are reliably safe and fair, these interventions must be specified in terms of norms that can be appropriately internalized by practitioners. 
The General Data Protection Regulation has had a mixed reception---while it did articulate normative landmarks for subfields to pay attention to, some of its requirements (e.g. consent as a legal basis for data processing) were highly underspecified. This specification vacuum empowered prominent private actors to advance their own standards in a way that is ethically questionable but politically effective, achieving market buy-in from enough other actors before the law can catch up \cite{utzInformed2019,nouwensDark}. Technical practitioners will need the ability to stand up and contest would-be standards publicly, rather than relying on the law to interpret systems before their sociotechnical scope has been appropriately modeled.
To avoid the solutionism trap, it is important to maintain a robust culture of questioning which problems should be addressed, and why these problems and not others: in the form of humility or a ``first, do no harm'' perspective.

\subsection{From Avoiding Traps to Anticipating Risks}
An important initial step for grappling with abstraction traps is for technical practitioners in fields of AI Safety, Fair ML, and HIL Autonomy to consider them explicitly when attempting to solve and formulate problems. 
In following with \cite{selbst2019fairness}, we find it most helpful to consider the traps in reverse order: is it worth designing a technical solution?
Can we adequately reason about how the technology will affect its social context?
Can the desired properties of the system be captured by mathematical abstractions?
Are the technical tools appropriate to the context?
And are all relevant actors included in the framing?
By considering these questions, researchers and practitioners will be encouraged to grapple with the plural temporalities defined by \textit{ongoing sociotechnical engagement} through the validation of assumptions behind featurization, optimization, and integration. 

While researchers in AI Safety, Fair ML, and HIL Autonomy are well positioned to begin asking these questions, it is only a first step. There is an inherent vulnerability in applying computational decision heuristics to vital social domains.
Autonomous AI systems introduce possibilities of catastrophic failure and normative incommensurability to contexts that were previously accessible only to human judgment and which we may never be able to exhaustively specify or completely understand. 
Beyond the mere \textit{avoidance of conceptual traps}, practitioners must learn to \textit{anticipate sociotechnical risks} as integral to the endeavor of building AI systems that interface with social reality.

The distinct intuitive approaches to risk taken by each of the examined subfields (\textit{extinction}, \textit{inequality}, and \textit{accident}) stem from alternative histories of the sorts of dangers faced when integrating systems within a normative social order. In other words, while these research communities have adopted tools and mathematical formalisms that purport to represent and work on discrete social phenomena, in fact the tools themselves are sociotechnical interventions, and their elaboration is justified according to historically-sedimented perceptions of risk.
Rather than systems that represent and affect specific social objects (e.g. people, institutions), we advocate for the concept of AI as a process of elaborating normative commitments whose technical refinement generates unprecedented positions \cite{dobbe2019hard}. From these positions, novel sociotechnical questions can be revealed, resolved, or deferred.

\section{Towards Clinical Training for Graduate Pedagogy}

How can researchers and practitioners learn to anticipate sociotechnical risks?
Awareness of abstraction traps may corroborate an appreciation of risks, but it does not provide the tools with which to anticipate or understand them. 
For example, pedagogical reforms based on coursework drawn from Science and Technology Studies, Philosophy, and Law may inspire a requisite caution in technical practitioners.
However, this caution remains insufficient to define the problem space of appropriate uses of AI. 
Instead, it will be necessary to encourage the coordination of technical and social scientists on these matters. In what follows, we interrogate this by evaluating possible reforms in graduate pedagogy.

Graduate students are a fruitful site for intervention for three primary reasons: 1) their educational role in shaping the next generation of engineers, 2) their role in pushing forward emerging areas of research and 3) their future as management and decision-makers at technology companies. Students should have the ability to recognize when a single development pipeline is trying to engage in multiple abstractions simultaneously because its metaphors are confused (e.g. the fact that certain AI Safety formalisms \cite{hadfield2016cooperative}, understood in terms of a principal-agent game, can function both as a form of mechanism design and as a kind of interface between user and robot).
It is further important that they have the ability to contest, merge, or even dissolve these frames if necessary. This will entail a major cultural transition in how the goals of graduate training are defined, moving away from failure-avoidance engineering in controlled environments to the responsible integration of technology in human contexts.

While there are efforts to widen the scope of a technical education and augment it with political and ethical training~ \cite{barabas2020studying,moore2020towards}, a truly sociotechnical graduate education would teach the skills of how to draw a technical bounding box as well as how to communicate those decisions to the publics that will have to reckon with the potential benefits and harms of new technology. Education cannot carve up the world into specific problem domains, but it could help coordinate concerns in a constructive manner that enables the development of context-appropriate validation metrics, as others have begun to do by synthesizing common technical pitfalls \cite{selbst2019fairness}.

While coursework lays the foundations for research, it cannot provide a descriptive ontology that would exhaustively capture sociotechnical risks in advance of active inquiry. 
Anticipating and mitigating such risks requires an immersion in the relevant social context, becoming richly familiar with its phenomenology from the human standpoint. Only by doing this is it possible to register the system specification in terms of the concrete normative stakes rather than abstract approximations of optimal behavior. 
This entails an ontological shift away from a purely mechanistic description of the domain in favor of a clinician's perspective, comparable in scope and significance to the emergence of modern medical and legal clinics \cite{bonner2000becoming,romano2016history,haydock1983clinical}. 
We believe a distinctly clinical approach to social problems---engaged and prolonged consultation, direct provision of service, relationships with clients, hands-on education overseen by professors---is the best approach.

Technical work will always rely on abstraction and framing to describe the environment in which a system is designed to function. It falls on technical researchers and practitioners to understand how to specify such a \emph{bounding box}, decide which frames and abstractions are valid and tractable as well as commensurate with stakeholder concerns, and articulate their choices to relevant communities with varying technical backgrounds. We see this ``clinician's eye,'' entailing effective framing and communication, as the most promising potential outcome of reforming AI pedagogy at the graduate level, and defer further investigation of clinical approaches in the context of CS education to future work.

\section{Conclusion}
The work of defining ``sociotechnical'' problems in AI development is ongoing. 
Systems themselves often make symbolic reference to situations, environments, or objects that are assumed to lie behind their representations unreflectively, allowing the same mathematical structures to propagate without interrogating key metaphorical frames. 
This norm results in a practice incommensurate with other expert professions’ standards of liability.
Along with the inconsistency between subfields, this makes it hard to define what constitutes an AI expert and how responsibility should be assigned when systems fail. 
Looking from the outside in, the legal and philosophical communities cannot enforce standards that are neither backed up by established forms of expertise 
nor immediately translatable outside the context of technical-mathematical formalism, meaning case law and abstract ethics cannot fully determine or guide sociotechnical regulation.

Given this normative indeterminacy, we argue there is no ready-made delineation of which technical tools are suited to which social problems, and instead look to prospective interventions nurturing new forms of inquiry into inherited notions of risk. 
On this view, interventions would embrace the notion that elaborating on sociotechnical problems and procedures is essential to the task itself, and practitioners would understand the sociotechnical simply as part of what they do. 
We argue this is the more sure path to effective norms for distinct subfields of AI development, and thus to the aims of Public Interest Technology.



%
\bibliographystyle{IEEEtran}
\bibliography{main}  

\end{document}